\documentclass[journal,comsoc]{IEEEtran}
\usepackage[utf8]{inputenc} 
\usepackage{cite}
\usepackage{amsfonts}
\usepackage{amsthm}
\usepackage{multirow}
\usepackage{footnote}
\usepackage{xcolor}
\usepackage{gensymb}
\usepackage{amsmath}
\usepackage{algorithmic}
\usepackage[pdftex]{graphicx}
\usepackage{url}
\usepackage{listings}
\usepackage{balance}

\makeatletter
\let\MYcaption\@makecaption
\makeatother
\usepackage[font=footnotesize]{subcaption}
\makeatletter
\let\@makecaption\MYcaption
\makeatother

\newcommand{\pkg}[1]{{\texttt{#1}}}

\begin{document}

\title{Design and Analysis of 5G Scenarios with \pkg{simmer}: An R Package for Fast DES Prototyping}

\author{
	Iñaki Ucar, José Alberto Hernández, Pablo Serrano, Arturo Azcorra 
	\thanks{The authors are with Universidad Carlos III de Madrid, Spain.}
	\thanks{Arturo Azcorra is also with IMDEA Networks Institute, Spain.}
}

\IEEEtitleabstractindextext{%
\begin{abstract}

Simulation frameworks are important tools for the analysis and design of communication networks and protocols, but they can result extremely costly and/or complex (for the case of very specialized tools), or too naive and lacking proper features and support (for the case of ad-hoc tools). In this paper, we present an analysis of three 5G scenarios using \pkg{simmer}, a recent R package for discrete-event simulation that sits between the above two paradigms. As our results show, it provides a simple yet very powerful syntax, supporting the efficient simulation of relatively complex scenarios at a low implementation cost. 

\end{abstract}

}

\maketitle
\IEEEdisplaynontitleabstractindextext
\IEEEpeerreviewmaketitle

\section{Introduction}

Simulation frameworks are undoubtedly one of the most important tools for the analysis and design of communication networks and protocols. Their applications are numerous, including the performance evaluation of existing or novel proposals, dimensioning of resources and capacity planning, or the validation of theoretical analyses, which are based on simplifying assumptions whose impact is to be assessed. 

In fact, simulation frameworks also make a number of simplifying assumptions, typically about components of the considered system that are not directly related to the performance variable of interest, to reduce complexity so the development of the scenario is easier and numerical figures are obtained faster. This ``complexity'' axis goes from very specialized, large simulation tools such as NS-3, OMNeT++, OPNET,\footnote{The list of network simulation tools is vast, see e.g. \url{http://people.idsia.ch/~andrea/sim/simnet.html}} to \emph{ad-hoc} simulation tools, consisting on hundreds of lines of code, typically used to validate a very specific part of the network or a given mathematical analysis. The latter are often developed over general-purpose languages such as C/C++ or Python, over numerical frameworks such as Matlab, or over some framework for discrete-event simulation.\footnote{See e.g. \url{https://en.wikipedia.org/wiki/List_of_discrete_event_simulation_software})}

On the one hand, the complexity of specialized tools (as their cost, if applicable) preclude their use for short-to-medium research projects, as the learning curve is typically steep plus they are difficult to extend, which is mandatory to test a novel functionality. On the other hand, the development of ad-hoc tools also require some investment of time and resources, lack a proper validation of their functionality and, furthermore, there is no code maintenance once the project is finished, for the few cases in which the code is made publicly available.

In this paper, we introduce the use of a recent event-driven simulation package, \pkg{simmer}, and show its applicability in fast prototyping three 5G-related scenarios. \pkg{simmer} sits between the above two complexity extremes and combines a number of features that supports, among others, versatility and repeatability. More specifically, some of the key advantages of \pkg{simmer} are as follows: (1)~it is based on the very popular R programming language, which benefits from a large community of users and contributors, but also natively supports the analysis of results via the many R statistical and visualization packages; (2)~furthermore, the code has been peer-reviewed \cite{simmer} and it is an official package, with numerous examples readily available, and potentially supported by a notable user population, and (3)~in addition to its ease of use and versatility, its code is partially optimized for speed, and therefore it can simulate relatively complex scenarios under reasonable times. 

We illustrate the use of \pkg{simmer} by simulating three different networking scenarios, which are inspired by current research trends regarding the design of the fifth-generation (5G) of mobile networks \cite{whatwillbe}. These diverse scenarios confirm the validity of \pkg{simmer} as a useful simulation tool that can support (at least as a first step) the dimensioning of communication systems, or can serve to quantify the trade-offs imposed by a given technology decision. More specifically, we consider the analysis of the following scenarios: 

\begin{itemize}
	\item Different design options for a \emph{crosshaul} scenario, where packetized in-phase and quadrature samples from fronthaul traffic are transmitted along backhaul data frames over the same links. Thanks to \pkg{simmer} and its support for statistical analysis, we can easily quantify delays under several queueing disciplines for different fronthaul/backhaul ratios, these being the metrics of interest for these scenarios.
	\item The impact of installing small cells in a fiber-to-the-premises scenarios. Here, we analyze two different approaches to support the highly-demanding cellular traffic along with the existing residential traffic, namely, the deployment of a remote radio head vs. the deployment of a small cell. 
	\item Massive Internet-of-Things scenarios, where thousands of metering devices share the same channel to upload their readings. Here, we analyze the impact of access parameters on performance, with a particular interest in the energy required to deliver the information, which will ultimately impact the lifetime of devices running on batteries. 
\end{itemize}

This article provides a quick overview of \pkg{simmer} and its key features. The analyses of the three considered 5G-related scenarios showcase the multiple benefits of different approaches and the versatility of \pkg{simmer} to easily implement their key features.


\section{An Introduction to \pkg{simmer}}

\pkg{simmer}\footnote{Available for download on the Comprehensive R Archive Network (CRAN), at \url{https://CRAN.R-project.org/package=simmer}} \cite{simmer} is a discrete event simulation (DES) library for the R Project\footnote{\url{https://www.R-project.org/}}, the open source programming language for statistical computing that has been receiving increased attention, primarily due to its widespread adoption for data science, analytics and statistical research. 

By developing \pkg{simmer} for R,  it can benefit from this growing ecosystem. Note that \pkg{simmer} does not aim at substituting NS-3 or OMNeT++, which are the \emph{de facto} standards for open-source network simulations. Instead, \pkg{simmer} is designed as a general-purpose DES framework with a human-friendly syntax and a very gentle learning curve. It can be used to complement other field-specific simulators as a rapid prototyping tool that enable insightful analysis of different designs. As we will illustrate in the next section, with \pkg{simmer} it is simple to simulate relatively complex scenarios, with the added benefit of the availability of many convenient data analysis and representation libraries, thanks to the use of R.

The R application programming interface (API) exposed by \pkg{simmer} revolves around the concept of \textit{trajectory}, which defines the ``path'' in the simulation for entities of the same type. A trajectory is a recipe for the arrivals attached to it, an ordered set of actions (or \textit{verbs}) chained together with the pipe operator (\texttt{\%>\%}, whose behaviour is similar to the command-line pipe). The following example illustrates a basic \pkg{simmer} workflow, modeling the classical case of customers being attended by a single clerk with infinite waiting space in a few lines of code: 

\begin{lstlisting}
library(simmer)

cust <- trajectory("customer") %>%
  seize("clerk", amount=1) %>%
  timeout(function() rexp(1, 2)) %>%
  release("clerk", amount=1)

env <- simmer("bank") %>%
  add_resource("clerk", capacity=1, queue_size=Inf) %>%
  add_generator("cust", cust, function() rexp(1, 1)) %>%
  run(until=1000)

arrivals  <- get_mon_arrivals(env)
resources <- get_mon_resources(env)
\end{lstlisting}

Given that both the time at the clerk and the time between customers are exponential random variables and infinite queue length, this example corresponds, in Kendall's notation, to an M/M/1 queue. It serves to illustrate the two main elements of \pkg{simmer}: the \texttt{trajectory} object and the \texttt{simmer} environment (or \textit{simulation environment}).

The \texttt{customer} trajectory (line 3) defines the behaviour of a generic customer: seize a clerk, spend some time, and release it. The \texttt{env} simulation environment (line 8) is then defined as one clerk with infinite queue size and a generator of customers, each one following the trajectory defined above. Based on this syntax, the flexibility is provided through a rich set of activities (more than 30) that can be appended to trajectories, which support: changing arrivals' properties (attributes, priority, batches), different interactions with the resources (select, seize, release, change their properties), and the generators (activate, deactivate, change their properties), and even the definition of branches (simple, depending on a condition, or parallel) and loops. Finally, some support to asynchronous programming is also provided (subscription to signals and registration of handlers).

Not only \pkg{simmer} provides a powerful yet simple syntax, but it is also \emph{fast}, for example, faster than equivalent frameworks such as SimPy and SimJulia for the Python and Julia languages respectively \cite{simmer}. The key for this speed is its underlying simulation core, which is written in C++. Furthermore, and perhaps more importantly, \pkg{simmer} implements automatic monitoring capabilities: every event is accounted for by default, both for arrivals (starting and ending times, activity time, ending condition, resources traversed) and resources (server and queue status), and all this information can be easily retrieved in standard R data frames for further processing of results (lines 13-14 of the \emph{clerk} example).

\begin{figure*}[t]
	\centering
	\begin{subfigure}[b]{.48\textwidth}
		\centering
		\includegraphics[width=\textwidth]{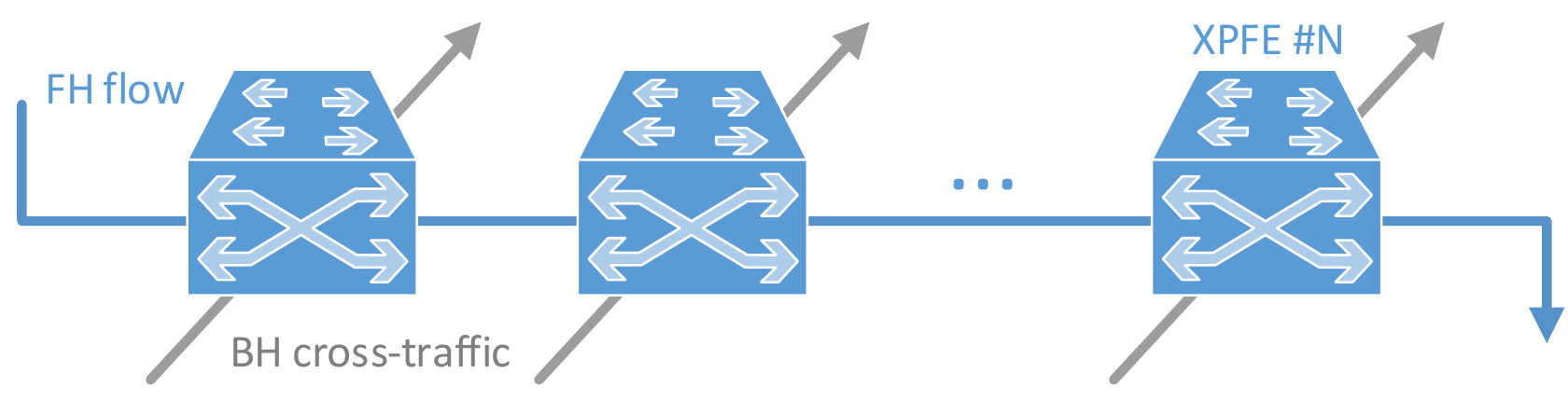}
		\caption{}
		\label{fig:scenario1def}
		\vspace{2ex}
		\includegraphics[width=.9\textwidth]{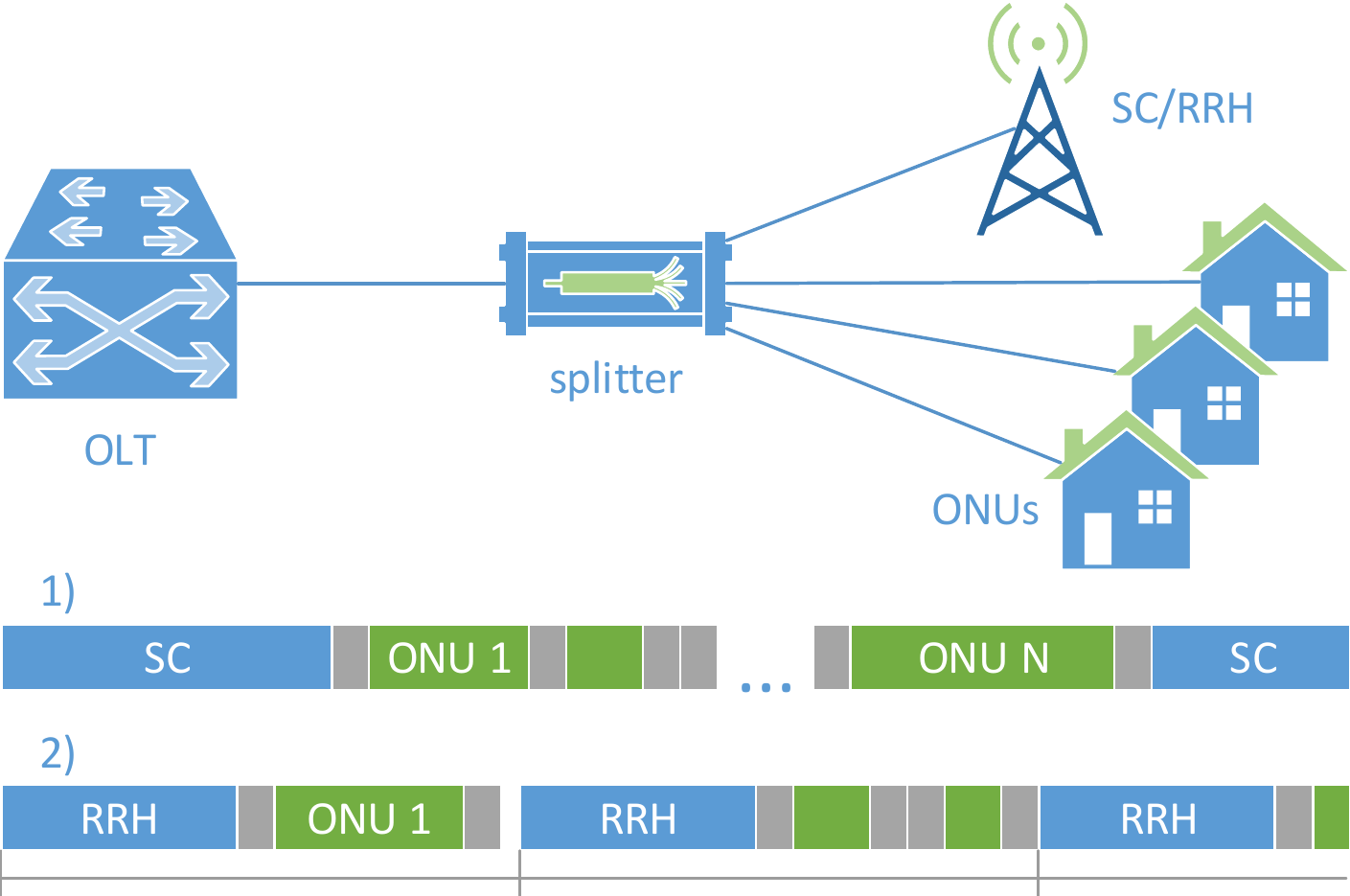}
		\caption{}
		\label{fig:scenario2def}
	\end{subfigure}
	\begin{subfigure}[b]{.48\textwidth}
		\centering
		\includegraphics[width=\textwidth]{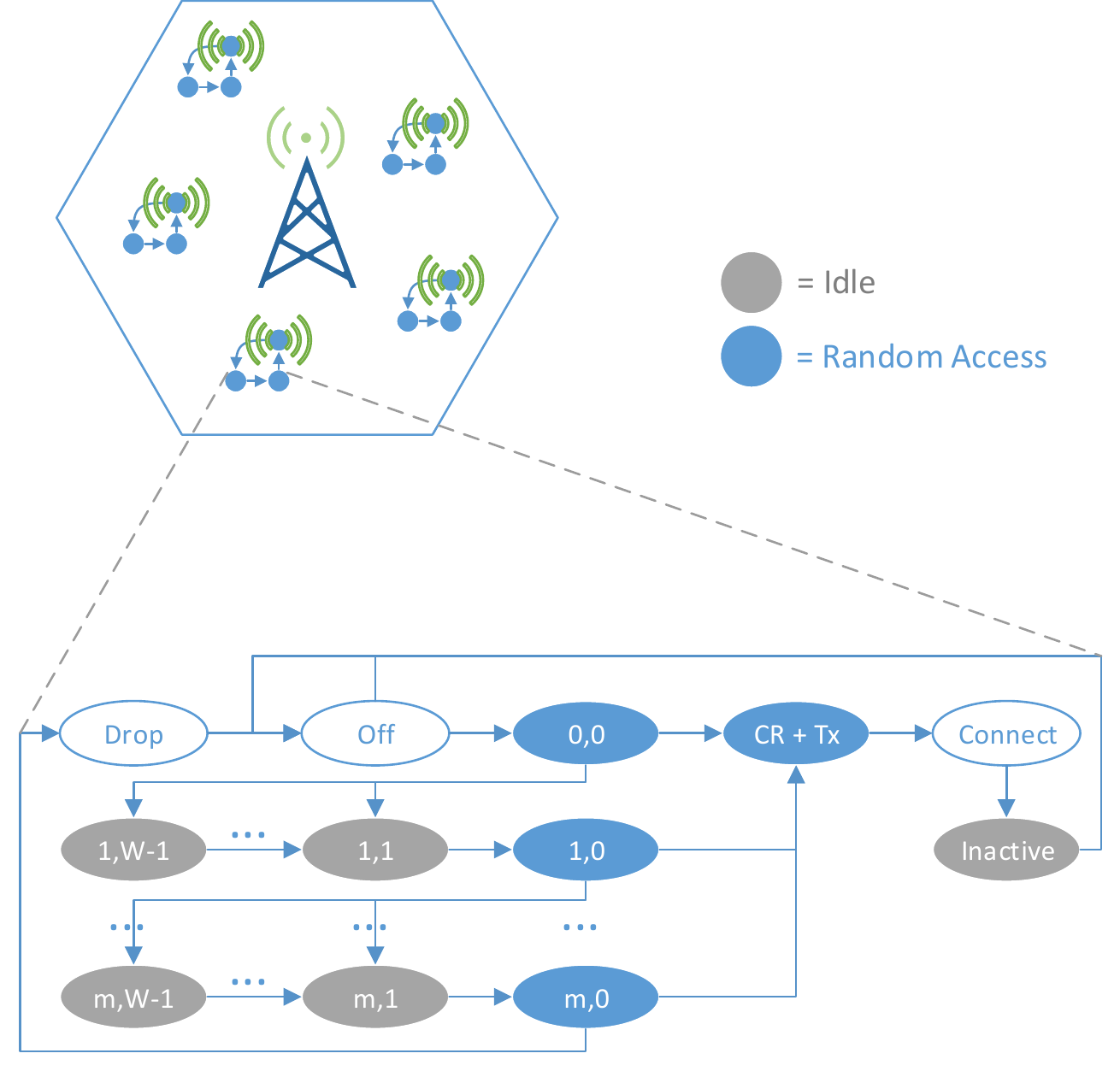}
		\caption{}
		\label{fig:scenario3def}
	\end{subfigure}
	\caption{Use cases: description of a) scenario no. 1; b) scenario no. 2; c) scenario no. 3.}
\end{figure*}


\section{Modeling 5G Scenarios}

In what follows, we will model and analyze three representative 5G scenarios using \pkg{simmer}. The source code for the use cases presented here, including configuration (definition of constants and parameters), simulation and analysis of results, is available online,\footnote{See the ``Articles'' section at \url{http://r-simmer.org}.} while some summary statistics of the simulations performed and their complexity are provided below.

\subsection{Crosshauling of FH and BH Traffic}

This scenario is motivated by the Cloud Radio Access Network (C-RAN) paradigm~\cite{overview}, where the mobile base station functionality is split into simple Remote Radio Heads (RRH), spread across the deployment and connected by fiber to centralized (and possibly virtualized) Base-Band Units (BBU), at the operators' premises. C-RAN is an architectural shift, aiming at providing CAPEX and OPEX savings while supporting better interference reduction and improved performance via Coordinated Multi-Point (CoMP).

In this C-RAN paradigm, fronthaul (FH) traffic from the RRU has stringent delay requirements, while backhaul (BH) traffic from the BBU has mild delay requirements. In a general topology, such as the one illustrated in Fig.~\ref{fig:scenario1def} (``scenario no. 1'') \cite{xhaul}, XPFEs (Crosshaul Packet Forwarding Element) will forward both types of traffic, and therefore introducing service differentiation might improve the ability to fulfil the delivery guarantees of FH traffic, which is in-line with the IEEE 802.1CM Time-Sensitive Networking for Fronthaul standard under development.\footnote{See \url{http://www.ieee802.org/1/pages/802.1cm.html}, Draft 0.6.} To quantify the benefits of this differentiation, we use \pkg{simmer} to simulate the scenario, this way supporting for example a design decision about the best scheme to deploy.

We assume that XPFEs are at 75\% load and operate at 40~Gb/s line rate. We assume 50\% of this load corresponds to FH traffic, and the other 50\% corresponds to BH traffic. Packet arrivals follow a Poisson process (although any other arrival process may be used in the simulation, for instance, bursty, self-similar); regarding packet sizes, FH packets are assumed to be CPRI (Common Public Radio Interface) Basic Frames of 80 byte length (i.e. CPRI option 4, see~\cite{own_cpri}), while BH packets follow the classical AMS-IX (Amsterdam Internet Exchange) trimodal function, namely 7 out of 12 packets are short (40 bytes), 4 out of 12 are medium size (576 bytes) and 1 out of 12 packets is long (1500 bytes).\footnote{Amsterdam Internet Exchange Ethernet Frame Size Distribution, Statistics available online at \url{https://ams-ix.net/technical/statistics/sflow-stats/frame-size-distribution}.}

We consider three different policies for service differentiation at the $N$ XPFEs: 

\begin{enumerate}
	\item No service differentiation.
	\item Strict Priority (SP) is given to FH over BH, but without preemption.
	\item SP is given to FH over BH with preemption.
\end{enumerate}

\begin{figure*}[t]
	\centering
	\begin{subfigure}[b]{0.48\textwidth}
		\includegraphics[width=\textwidth]{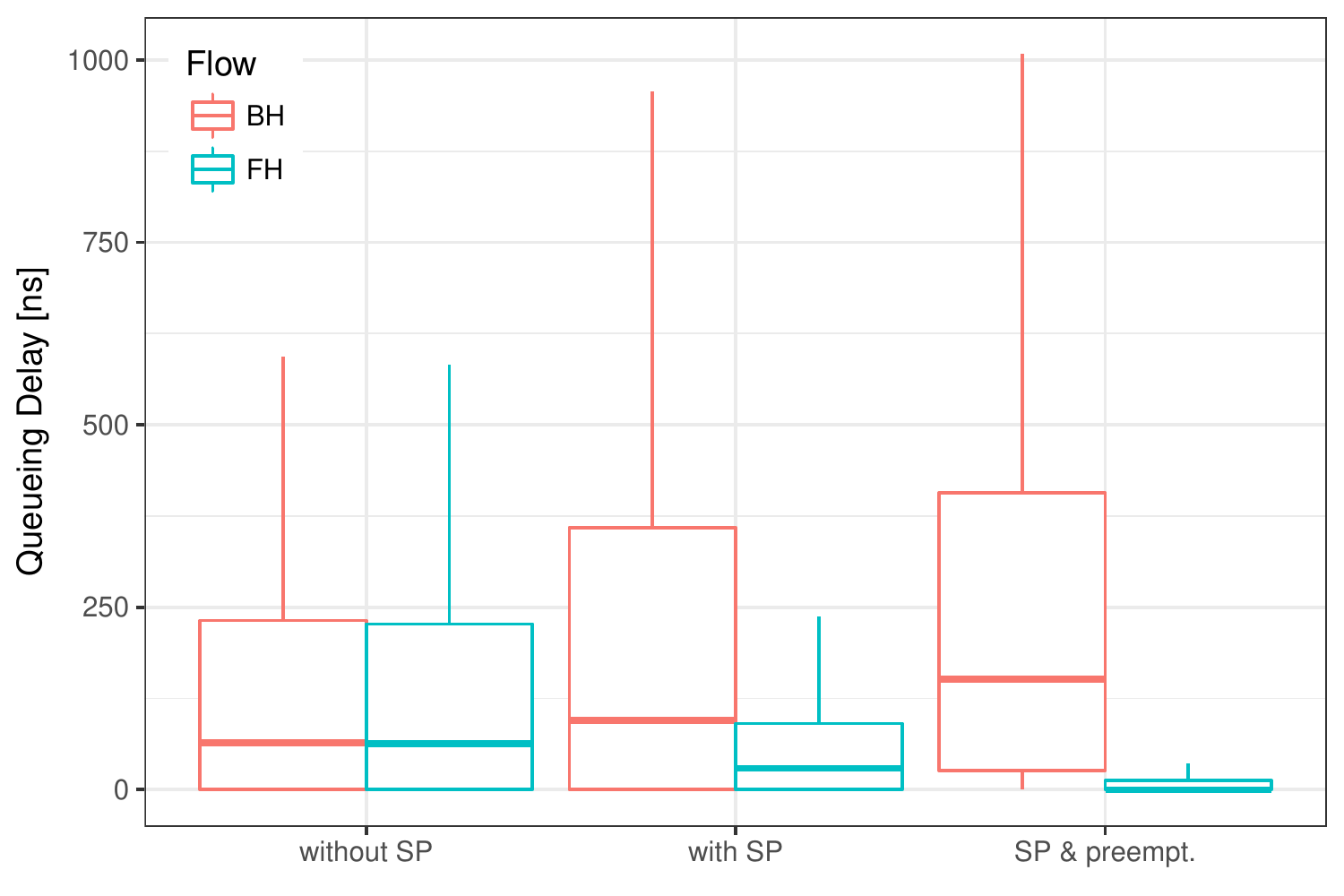}
		\caption{}
		\label{fig:scenario1a}
	\end{subfigure}
	\begin{subfigure}[b]{0.48\textwidth}
		\includegraphics[width=\textwidth]{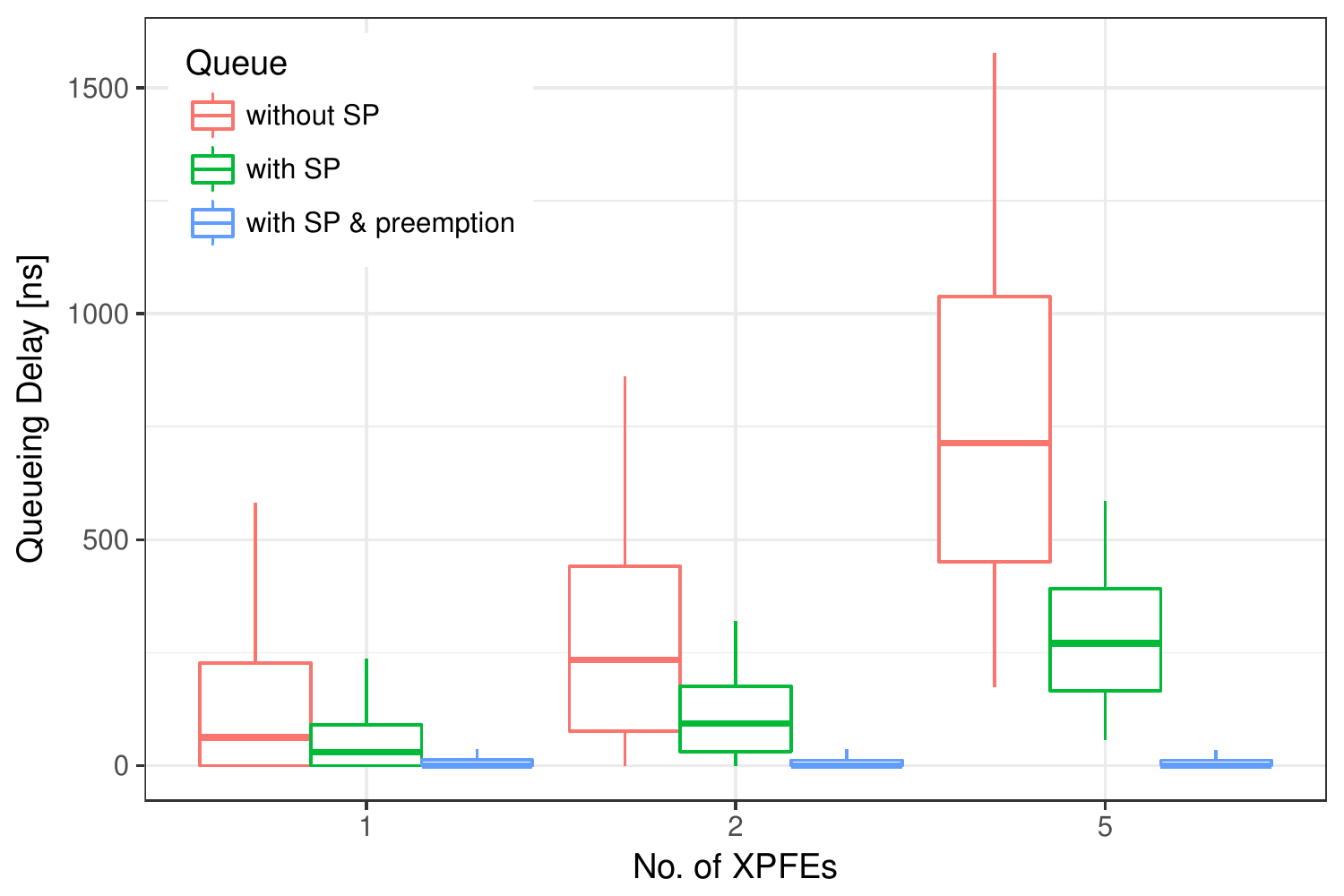}
		\caption{}
		\label{fig:scenario1b}
	\end{subfigure}
	\caption{Queueing delay experienced by FH and BH traffic under different QoS policies: a) FH/BH queueing delay comparison for a single XPFE; b) accumulated FH delay for several XPFEs. The whiskers represent the 5th and 95th percentiles.}
	\label{fig:scenario1}
\end{figure*}

For each of these policies, we first simulate a one-XPFE scenario with the traffic characteristics described above, and compute the queueing delay for FH and BH traffic. The results are depicted in Fig.~\ref{fig:scenario1a}, where we use box-plots to illustrate the \{5, 25, 50, 75, 95\}-th percentiles. 

As the figure shows, the first strategy (`without SP') results in both traffic experiencing the same queueing delay. The second strategy (`with SP'), that is, service differentiation without preemption, results in a much improved service for FH traffic, as FH packets only have to wait for other FH packets ahead and, sometimes, one BH packet being served (i.e., the residual service time). The third strategy (`with SP \& preemption') implements a preemption strategy, discarding BH traffic from the transmission line if a FH packet arrives which, as the figure shows, does significantly improve delay performance (the 95-th percentile drastically decreases), which was caused by the long transmission times of long BH frames (1500 bytes). For this scenario, we conclude that a preemption strategy reduces queuing delay to the minimum, with very high delivery guarantees.

We next analyze scenarios where the FH has to transverse $N$ XPFEs in tandem, each one also serving BH traffic, as illustrated in Fig.~\ref{fig:scenario1def}. We depict in Fig.~\ref{fig:scenario1b} the queueing delays of FH traffic for $N=\{1, 2, 5\}$, under the three considered policies (the $N=1$ case corresponds to the same results as in the previous experiments). As shown, FH queueing delay accumulates after traversing multiple XPFEs for the first and second policies, both in terms of median and percentiles, hence jitter too. Only the SP strategy with preemption keeps both delay and jitter values extremely low, since FH packets have only to wait if the server is occupied by other FH packets. In contrast, the SP strategy without preemption shows accumulated packet delay and packet delay variability after traversing multiple XPFEs. 

\subsubsection*{Implementation Details}

The simulation implements a single trajectory for the FH traffic, \texttt{fh\_traffic}, that seizes the $N$ XPFEs sequentially. Additionally, a list of $N$ trajectories called \texttt{bh\_traffic}, one for each XPFE, models the interfering BH traffic. The XPFEs are defined as resources with \texttt{capacity=1} and infinite queue length. A generator of FH traffic plus $N$ generators of BH traffic are attached to their respective trajectories. All the elements are encapsulated into a function and a number of constants are parametrized, namely, the number of XPFEs, FH traffic's priority and whether the XPFEs should be preemptive or not. As can be seen, all the \texttt{cases}, defined as a data frame (one case per row), are easily parallelized using standard R tools (i.e., the R-core \pkg{parallel} package).

All the monitoring information automatically collected by \pkg{simmer} for the arrivals (BH and FH packets) can be retrieved as a data frame using the \texttt{get\_mon\_arrivals()} method. This enables further analysis and visualization in only a few lines of code using the \pkg{dplyr}\footnote{\url{https://CRAN.R-project.org/package=dplyr}} and \pkg{ggplot}\footnote{\url{https://CRAN.R-project.org/package=ggplot2}} packages.

\subsection{Mobile Traffic Backhauling with FTTx}

We next consider the case of a residencial area with a Fiber-To-The-Premises (FTTx) infrastructure, that is, an Optical Distribution Network (ODN), composed of the Optical Line Terminal (OLT), splitters, and the Optical Network Unit (ONU) at the users' premises. As Fig.~\ref{fig:scenario2def} illustrates (``scenario no. 2''), we assume that an operator is planning to deploy an antenna, carrying the mobile traffic over the ODN, and is considering two implementation options:

\begin{enumerate}
	\item Deployment of a Small Cell, reducing the amount and requirements of the generated traffic.
	\item Deployment of an RRH, following the C-RAN paradigm discussed above, which would therefore generate time-sensitive FH traffic. 
\end{enumerate}

In both cases, we analyze the upstream channel of a Time-Division Multiplexed Passive Optical Network (TDM-PON) providing broadband access to the residential users and the mobile users. We assume for simplicity the case of ITU-T G.984 Gigabit PON (it would be straightforward to extend the results to other TDM-based PONs, such as XG-PON, XGS-PON, EPON, 10G-EPON), with a total capacity of 1.25~Gb/s in the upstream, and where each ONU generates 20~Mb/s of upstream traffic. 

\subsubsection{Small Cell}

Here we assume that the small cell generates 150~Mb/s peak traffic, and shares the ODN with 31 residential users, which corresponds to an average total load of 61.6\%. We assume bursty arrivals following a compound Poisson model, where both the bursts and the burst length are Poisson-distributed, with a mean of 20~packets. As in the previous use case, packet sizes are randomly chosen following the AMS-IX trimodal distribution, yielding an average burst size of 6407~Bytes. In this case, we assume a Dynamic Bandwidth Allocation (DBA) algorithm similar to the IPACT protocol\cite{kramer}, where each ONU requests at the end of its transmission window resources for the next cycle, while the OLT receives such requests and decides when and for how long each ONU may transmit its data in the next cycle granting transmission windows in a round-robin fashion. We assume a $1~\mu s$ guard-time between transmission windows of consecutive ONUs.

We consider four different quality-of-service (QoS) policies for the DBA, these being defined by the maximum supported request per cycle per ONU (the cellular traffic is always granted their requests in each case). If an ONU requests more that this maximum transmission window (namely, 1500~B, 3000~B, 6000~B or infinite, which means no limit), the OLT will only grant this maximum, and therefore the rest of the traffic will queue at the ONU. 

\subsubsection{RRH Backhauling}

Next, we assume an RRH generating FH traffic, following the MAC-PHY (Media Access Control, physical layer) functional split, that is, one OFDM (Orthogonal Frequency Division Multiplexing) symbol is sampled, quantized (approx. 6000~B) and transmitted to the BBU every $66.67~\mu$s, this resulting in an approximate rate of 720~Mb/s. This FH traffic shares the ODN with now 7 residential ONUs, this resulting in an approx. total load of 68.8\%. To fulfil the tight delivery requirements of FH traffic, we assume that the DBA algorithm guarantees periodic TDM reservations for the FH traffic, and then the ONUs use a similar algorithm as before to share the rest of the available bandwidth. 

\begin{figure}[t]
	\centering
	\includegraphics[width=0.48\textwidth]{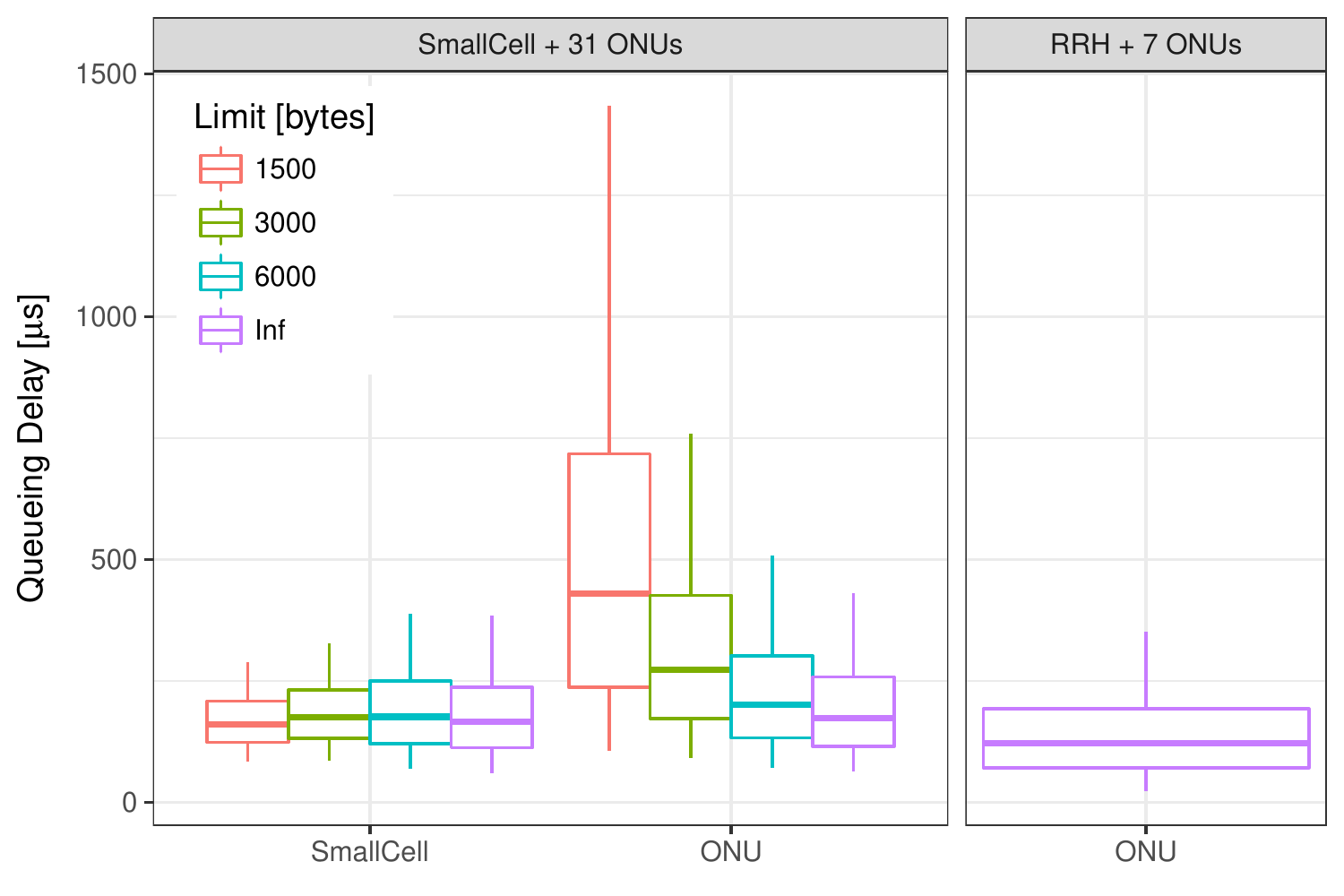}
	\caption{Upstream delay for small cell and residential users sharing the upstream channel in a TDM-PON.}
	\label{fig:tdm-pon-delay}
\end{figure}

We depict in Fig.~\ref{fig:tdm-pon-delay}~(left) the results corresponding to the uplink delay in Case 1, that is, one Small Cell and 31 ONUs, for both types of traffic and the four QoS policies. Like in the previous case, we focus on the queueing delay, and represent the different percentiles with box-and-whisker plots. As the figure shows, when requests are not limited (`Inf') the traffic from the ONUs and the Small Cell experience the same delay, as expected. However, the moment the DBA algorithm enforces a limit to ONUs requests, differences appear: the Small Cell delays are smaller and less disperse, while the ONU results are longer and show more variability. It can be seen that with a 1500~B limit, the delivery guarantees for the Small Cell traffic are very tight (e.g., 95-th percentile around 250~$\mu$s), which is achieved at the expense of significant delays for the ONUs. 

Next, we depict in Fig.~\ref{fig:tdm-pon-delay}~(right) the results for the uplink delay in Case 2, that is, a RRH and 7 ONUs. We do not depict here the delay results corresponding to the FH traffic, as in this case the reservation mechanism guarantees its delivery with zero queueing delay. As the figure show, the delay performance obtained by users is slightly better than in the previous case, although less users can be accommodated due to traffic demands. 

\subsubsection*{Implementation Details}

This scenario serves to illustrate a different strategy to code a simulation with \pkg{simmer}. Instead of a trajectory attached to an unlimited generator of arrivals, the \texttt{OLT} is defined here as a trajectory with a single worker in an infinite loop by using the \texttt{rollback()} activity. This \texttt{OLT} executes the DBA logic encapsulated into the \texttt{set\_next\_window()} function, and timely activates the ONUs, defined as resources, in a round-robin fashion. This is achieved by modifying the capacity of a given ONU resource to meet the number of packets allocated for the next transmission window.

$N$ generators feed traffic into $N$ trajectories defined as a list of \texttt{ONUs}. Packets arriving there, similarly to the previous use case, first seize the corresponding ONU resource (which acts as a \textit{token bucket}) and, as soon as the \texttt{OLT} increments the capacity, they seize the link to be transmitted. An additional ONU is defined with a different traffic rate for the small cell case. As for the RRH case, the \texttt{RRH} trajectory is added, which holds a single worker in a loop seizing and releasing the link periodically.

As in the previous use case, a number of constants are parametrized (scenario, number of ONUs and TDM reservation limit) and all the combinations are parallelized with the R-core \pkg{parallel} package. Finally, the monitoring statistics for the arrivals are retrieved, and a similar analysis is made using \pkg{dplyr} and \pkg{ggplot} in very few lines of code.

\subsection{Energy Efficiency for Massive IoT}

Finally, we consider the case of a massive Internet-of-Things (mIoT) scenario, a use case for Long Term Evolution (LTE) and next-generation 5G networks, as defined by the Third Generation Partnership Project (3GPP) in \cite{miot}. As Fig.~\ref{fig:scenario3def} (top) illustrates, we consider a single LTE macrocell in a dense urban area. The buildings in the cell area are populated with $N$ smart meters (for electricity, gas and water), and each meter operates independently as a Narrowband IoT (NB-IoT) device.

\begin{figure}[t]
	\centering
	\includegraphics[width=0.48\textwidth]{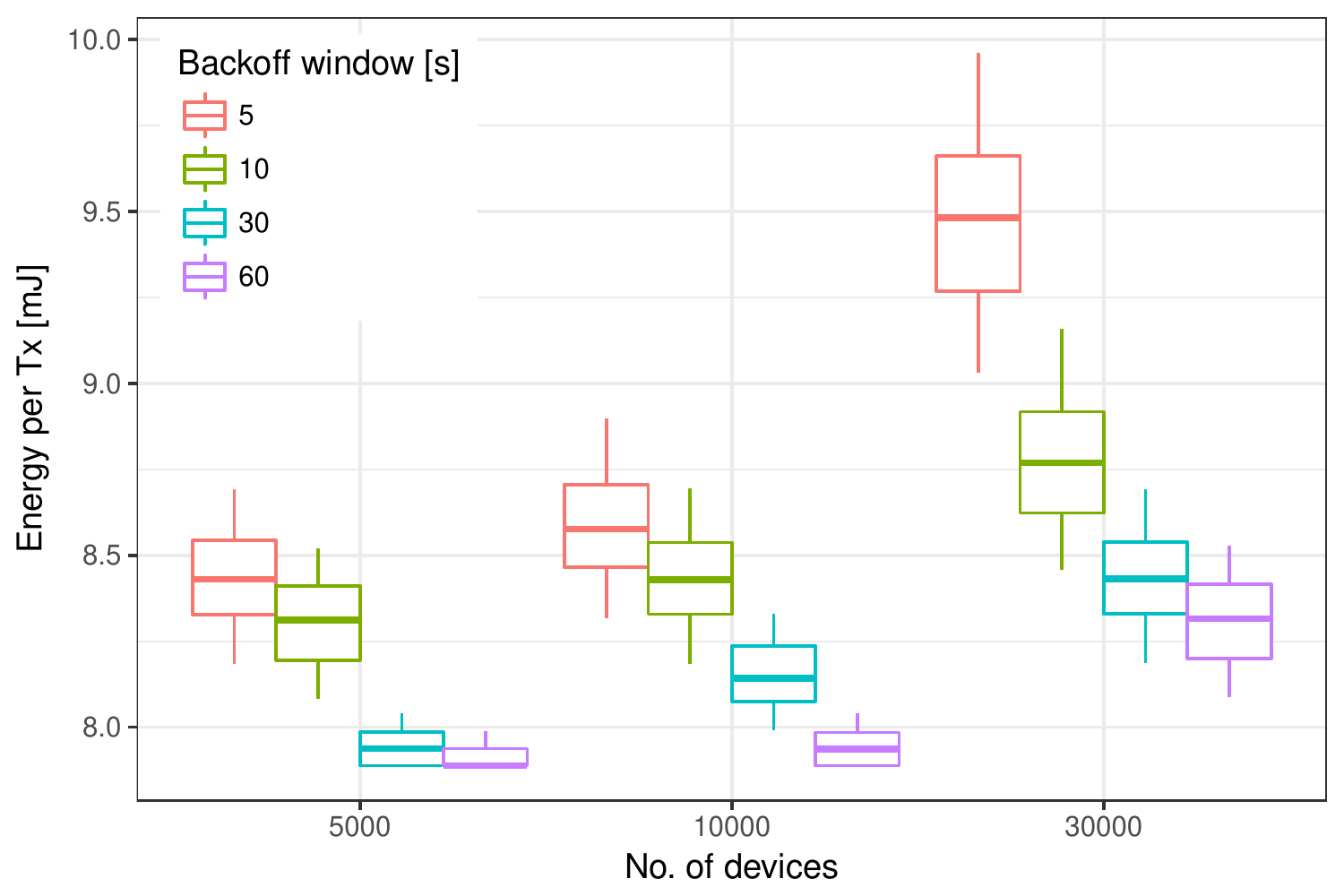}
	\caption{Energy consumption per transmission attempt for different traffic models and number of devices.}
	\label{fig:iot-energy}
\end{figure}

The devices' behaviour is modeled following the diagram depicted in Fig.~\ref{fig:scenario3def} (bottom), which is a simplified version of the Markov chain model developed in \cite{iotModel} (Fig.~5). A device may stay in \texttt{RRC Idle} (`Off') and awakes with some periodicity to upload its reading. This communication phase encompasses a contention-based random access (RA) procedure, with a backoff time randomly chosen between $(0,W)$ time slots and up to $m$ retransmissions. If the connection request fails, the reading is dropped and the device returns to the `Off' state. If the connection is successful, we assume that the device implements the Control Plane Cellular IoT (CP) optimization \cite{iotModel}, so that the data is transmitted over the \texttt{RRC Connection} request phase using the Non Access Stratum (NAS) level. Then, the device has to wait (`Inactive') until the connection is released, and eventually returns to the `Off' state.

The goal of this use case is to study the effect of synchronization across IoT devices (for instance, due to a power outage) in the energy consumption. As in \cite{iotCollision}, we assume that a device provides its readings as often as every hour, and the cases of $N=\{5, 10, 30\} \cdot 10^3$ devices in one cell are considered. In order to study different levels of synchronization, each node implements an additional backoff window prior to the RA procedure. Furthermore, we selected $m=9$ and $W=20$; the rest of the parameters (power consumption, timings, message sizes...) can be found in \cite{iotModel} (Table~I).

Fig.~\ref{fig:iot-energy} shows the results of the simulation for one day. It depicts the energy consumed per reading considering a uniform backoff window between $0$ and 5 (\textit{highly synchronized}), 10, 30 and 60 seconds (\textit{non-synchronized}). As the number of devices and the level of synchronization grow, the random-access opportunities (RAOs) per second grow as well producing more and more collisions. These collisions cause retries and a noticeable impact in the energy consumption (up to 12\% more energy per reading). Therefore, this use case shows the paramount importance of randomizing node activation in mIoT scenarios in order to avoid RAO peaks and a premature battery drain.

\subsubsection*{Implementation Details}

This scenario requires a single \texttt{meter} trajectory implementing the logic of each IoT device in an infinite loop, and $N$ workers are attached to it at $t=0$. Each device registers itself for a given signal (``reading'') and waits in sleep mode until a new reading is requested, which is triggered by a secondary trajectory (\texttt{trigger}). As soon as a new reading is signalled, the RA procedure starts by randomly selecting one of the 54 preambles available, which are defined as resources. The process of seizing a preamble encompasses two sub-trajectories:

\begin{itemize}
	\item If there are no collisions, the preamble is successfully seized and the \texttt{post.seize} sub-trajectory is executed, which transmits a reading.
	\item If there is collision, rejection occurs, and the \texttt{reject} sub-trajectory is executed, which performs the RA backoff (for a random number of slots) and restarts the RA procedure (for a maximum of $m$ retries).
\end{itemize}

Both sub-trajectories set the appropriate power levels $P$ for the appropriate amount of time. In this case, these power levels throughout the simulation time are retrieved with the \texttt{get\_mon\_attributes()} method. Again, the energy is concisely computed and represented using \pkg{dplyr} and \pkg{ggplot} packages.

\subsection{Wrap up}

Thanks to these scenarios, we have demonstrated the usability and suitability of \pkg{simmer} for fast prototyping of three different 5G scenarios. The code developed highlights some of the characteristics that make \pkg{simmer} attractive for researchers and practitioners in communications research: 

\begin{itemize}
  \item A novel and intuitive trajectory-based approach that simplifies the simulation of large networks of queues, including those with feedback.
  \item Flexible resources, with dynamic capacity and queue size, priority queueing and preemption.
  \item Flexible generators of arrivals that can draw interarrival times from any theoretical or empirical distribution via a function call.
  \item Asynchronous programming features and monitoring capabilities, which helps the researcher focus into the model design.
\end{itemize}

\begin{table}[t]
	\centering
	\caption{Overview of simulation features.}
	\begin{tabular}{r|rrr}
		\hline
		 							& Use case 1 	& Use case 2 	& Use case 3 	\\
		\hline\hline
		Simulation time (s)			& $\sim$10 		& $\sim$150 		& $\sim$150 		\\
		No. of parallel scenarios 	& 12 			& 5 				& 12 			\\
		Max. events, 1 scenario		& 2 215 076 		& 4 427 839 		& 28 364 172		\\
		Total no. of events 			& 14 565 424		& 11 196 337		& 98 952 165		\\
		Implementation lines 		& 30				& 97				& 42				\\
		Analysis + plotting lines 	& 28				& 18				& 14				\\
		\hline
	\end{tabular}
	\label{tab:overview}
\end{table}

Table~\ref{tab:overview} summarizes the main simulation statistics for each scenario (simulation time is computed with a machine equipped with an Intel(R) Xeon(R) CPU E5-2620 v4 @ 2.10GHz x4 (32 cores) and 64 GB of RAM, Debian GNU/Linux 8, R 3.3.2 and \pkg{simmer} 3.6). These numbers attest that \pkg{simmer} can be used to simulate relatively complex scenarios with very few lines of code (under 100 lines in all cases). Furthermore, the automatic monitoring capabilities embedded in \pkg{simmer}, and the integration with the R language, enable sophisticated analyses and visualizations just with a few more lines. It is likewise remarkable the ease with which multiple scenarios, with different parameters, can be simulated concurrently thanks to base R functions. Thus, exploring a large number of combinations of parameter values is not only straightforward, but also as fast as the slowest thread given enough number of CPU cores available.


\section{Conclusions}

In this paper, we have introduced the use of the \pkg{simmer} R package for the simulation of communication scenarios.  We have illustrated its simple yet powerful syntax, and have demonstrated its ease of use and functionality with the analysis of three 5G-inspired scenarios, corresponding to radio, access and metro deployments. The results obtained, which can be easily computed thanks to the powerful capabilities of R, help taking design decisions related to hardware choices, traffic prioritization or access scheme configuration. Because of these, we believe \pkg{simmer} is a powerful tool to validate analytical studies, or to complement the use of more complex and costly simulation frameworks.


\section*{Acknowledgements}

This article has been partially supported by the 5G-City project (TEC2016-76795-C6-3-R) and the TEXEO project (TEC2016-80339-R), both funded by the Spanish Ministry of Economy and Competitiveness.



\section*{Biographies}

\begin{IEEEbiographynophoto}{Iñaki Ucar} (inaki.ucar@uc3m.es) received his M.Sc.Eng. in Telecommunications Engineering and M.Sc. in Communications from Universidad Pública de Navarra (UPNA) in 2011 and 2013 respectively, and his M.Sc. in Telematics Engineering from Universidad Carlos III de Madrid (UC3M) in 2014. Currently, he holds the position of Teaching Assistant and pursues his Ph.D. in the Department of Telematics Engineering of UC3M. His work focuses on energy efficiency of wireless networks.
\end{IEEEbiographynophoto}

\begin{IEEEbiographynophoto}{José Alberto Hernández} (jahgutie@it.uc3m.es)
completed his five-year degree in Telecommunications Engineering at UC3M in 2002, and his Ph.D. degree in Computer Science at Loughborough University, Leicester, United Kingdom, in 2005. He has been a senior lecturer in the Department of Telematics Engineering since 2010, where he combines teaching and research in the areas of optical WDM networks, next-generation access networks, metro Ethernet, energy efficiency, and hybrid optical-wireless technologies. He has published more than 75 articles in both journals and conference proceedings on these topics. He is a co-author of the book \emph{Probabilistic Modes for Computer Networks: Tools and Solved Problems}.
\end{IEEEbiographynophoto}

\begin{IEEEbiographynophoto}{Pablo Serrano} (pablo@it.uc3m.es) 
received his degree in Telecommunications Engineering and his Ph.D.~from Universidad Carlos III de Madrid (UC3M) in 2002 and 2006, respectively. He has been with the Telematics Department of UC3M since 2002, where he currently holds the position of Associate Professor. He has over 70 scientific papers in peer-reviewed international journal and conferences. He has served as guest editor for Computer Networks, and on the TPC of a number of conferences and workshops including IEEE INFOCOM, IEEE WoWMoM and IEEE Globecom.
\end{IEEEbiographynophoto}

\begin{IEEEbiographynophoto}{Arturo Azcorra} (\texttt{azcorra@it.uc3m.es})
received his M.Sc. degree in Telecommunications Engineering from the Universidad Polit\'ecnica de Madrid in 1986 and his Ph.D. from the same university in 1989. In 1993, he obtained an M.B.A. from the Instituto de Empresa. He has a double appointment as Full Professor (with chair) at the Telematics Engineering Department of University Carlos III of Madrid and as Director of IMDEA Networks. Prof.~Azcorra has coordinated the CONTENT and E-NEXT European Networks of Excellence, has served as a Program Committee Member in numerous international conferences and has published over 100 scientific papers in books, international journals and conferences.
\end{IEEEbiographynophoto}

\balance

\end{document}